\documentclass[reprint,amsmath,amssymb,aps]{revtex4-2}
\usepackage{mathtools}
\usepackage{hyperref}
\usepackage{physics}
\usepackage{graphicx}
\begin{document}

\title{Strong-coupling mechanism of the pseudogap in small Hubbard clusters}
\author{Edwin W. Huang}
\email{edwinwhuang@gmail.com}
\affiliation{Department of Physics and Institute for Condensed Matter Theory, University of Illinois at Urbana-Champaign, Urbana, Illinois 61801, USA}
\begin{abstract}
In the hole-doped cuprates, the pseudogap refers to a suppression of the density of states at low energies, in the absence of superconducting long-range order. Numerous calculations of the Hubbard model show a pseudogap in the single-particle spectra, with striking similarities to photoemission and tunneling experiments on cuprates. However, no clear mechanism has been established. Here, we solve the Hubbard model on $2\times2$ clusters by exact diagonalization, with integration over twisted boundary conditions. A pseudogap is found in the single-particle density of states with the following characteristics: a decreasing energy scale and onset temperature for increased hole-doping, closure at a critical hole doping near 15\%, absence upon electron-doping, particle-hole asymmetry indicated by the location of the gap center, and persistence in the strong-coupling limit of $U/t \to \infty$. Studying the many-body excitation spectrum reveals that the pseudogap in single-particle spectra is due to orthogonality between bare electrons and the lowest energy excitations for $U/t \gtrsim 8$.
\end{abstract}
\maketitle
\emph{Introduction.---}In discussions of the hole-doped cuprates, the pseudogap is defined as the suppression of electronic density of states at low energies that is present well above the superconducting transition temperature \cite{keimer2015}. The pseudogap is manifest most directly in probes of the single-particle Green's function, such as photoemission and tunneling spectroscopy. There, it has been demonstrated to affect a substantial portion of the phase diagram up to a characteristic temperature $T^*$, which is a decreasing function of doping and vanishes above a critical doping $p^*$ that is slightly above optimal doping. The phase diagram below $T^*$ is particularly notable for the variety of complex phenomena and intertwined orders present \cite{keimer2015,fradkin2015}. Many scenarios have been proposed to explain the pseudogap, and determining its mechanism is a central question in the study of high-temperature superconductivity in cuprates.

Calculations of the 2D Hubbard model have found a pseudogap in the single-particle spectra, with properties closely resembling those of photoemission and tunneling spectra of cuprates \cite{prelovsek1999,maier2000,maier2001,jarrell2001,haule2002,haule2003,stanescu2003,senechal2004,sadovskii2005,stanescu2006,tremblay2006,kyung2006,macridin2006,gull2008,liebsch2009,vidhyadhiraja2009,ferrero2009,sakai2009,ferrero2009epl,sordi2012,sordi2012srep,gull2013,kohno2014,gunnarsson2015,yang2016,chen2017,wu2017,wu2018,braganca2018,kuzmin2020}. These include its extent in the temperature-doping phase diagram \cite{prelovsek1999,jarrell2001,vidhyadhiraja2009,sordi2012,sordi2012srep,kuzmin2020} and the contrasting behaviors in the anti-nodal and nodal regions of the Brillouin zone, leading to the phenomena of Fermi arcs \cite{stanescu2006,kyung2006,macridin2006,sakai2009,ferrero2009epl,kohno2014,braganca2018,kuzmin2020}. Apart from single-particle properties, other important aspects of pseudogap phenomenology in cuprate have been found and studied in the Hubbard model, including suppression of the spin susceptibility \cite{moreo1993,jarrell2001,bonca2003,kokalj2017,chen2017,reymbaut2019} and presence of spin and charge density waves \cite{white2003,zheng2017,huang2018}. While details of these calculations vary, a strong empirical argument can be made that the Hubbard model captures much of the essential physics relevant to the pseudogap in the cuprates.

In spite of these successes of the Hubbard model in accounting for experimental phenomenology, the cause or mechanism of the pseudogap remains an open problem. Even strictly within the confines of calculations of the Hubbard model and the closely related $t$-$J$ model, no consensus has been reached among the various proposed explanations of the pseudogap. Nevertheless, important clues may be gleaned from these calculations. First, most of these calculations involve solving a small cluster that is part of the infinite system, via a generalization of dynamical mean-field theory or by cluster perturbation theory (CPT). Some of these, such as the CPT calculations in \cite{senechal2004,kohno2014,yang2016,kuzmin2020}, do not permit long-range order, implying that short-range correlations are sufficient for generating the pseudogap. Viewed as simulations of the Hubbard model, such calculations are incorrect in the sense that they do not produce low-temperature ordered phases. However, it is plausible that they provide an accurate account of spectra down to energies relevant to the pseudogap. Second, calculations studying the dependence of the pseudogap on the interaction strength $U/t$ have demonstrated its persistence in the large $U/t$ limit \cite{senechal2004,tremblay2006,kyung2006}. The size of the gap is found to be $\mathcal{O}(t)$ rather than $\mathcal{O}\qty(\frac{t^2}{U})$. (This is not to say that the size of the gap is necessarily closer to $t$ than to $J=4\frac{t^2}{U}$, but that the size of the gap is not dependent on $U$ in the limit $U/t \to \infty$.) It is therefore not clear whether mechanisms of the pseudogap based on scattering of electrons off antiferromagnetic spin fluctuations, as often proposed in the context of weak- and intermediate-coupling calculations, are applicable in the strong-coupling limit. 

\emph{Methodology.---}Here, our approach is to rely on exact diagonalization calculations of small Hubbard model clusters. It is not the goal to solve the Hubbard model in the thermodynamic limit, but to determine whether the essential physics of the pseudogap is present in small clusters amenable to exact calculations. An analogy with the physics of the Mott gap is instructive to motivate this approach. Solving the Hubbard model on a single site is sufficient not only to obtain the Mott gap, but also to demonstrate its mechanism. Though this has limitations in describing the Mott transition and the antiferromagnetic ground state, the essence of the Mott gap and the associated upper and lower Hubbard bands are properly captured.

\begin{figure}[t]
\includegraphics{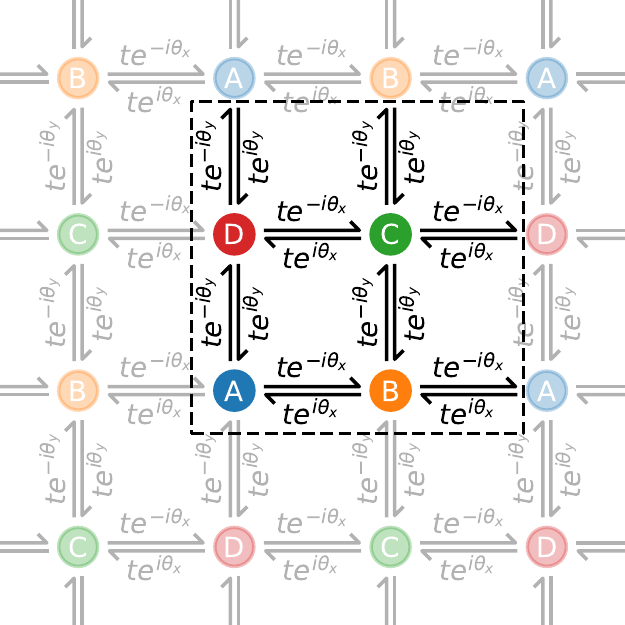}
\caption{Nearest neighbor hopping matrix elements for a $2\times2$ cluster with twisted boundary conditions. The next-nearest neighbor hopping $t'$ is not shown here for clarity.}
\label{fig:1}
\end{figure}

Specifically, we consider $2\times2$ Hubbard model clusters with twisted boundary conditions. We choose the gauge where the fermion operators are periodic ($c_{i\sigma}^\dagger \equiv c_{i+L_x \hat{x},\sigma}^\dagger \equiv c_{i+L_y \hat{y},\sigma}^\dagger$) and implement the twist as a phase in every hopping matrix element, as illustrated in Fig.~\ref{fig:1} for nearest neighbors. The parameter $\boldsymbol{\theta}$ parametrizes the twist; for instance, antiperiodic boundary conditions correspond to $\boldsymbol{\theta} = (\frac{\pi}{L_x}, \frac{\pi}{L_y})$. The Hamiltonian for a given twist $\boldsymbol{\theta}$ is given by
\begin{equation}
H_{\boldsymbol{\theta}} = - \sum_{i j \sigma} t_{i j} e^{-i \boldsymbol{\theta} \cdot (\mathbf{r}_i - \mathbf{r}_j)} c_{i \sigma}^\dagger c_{j \sigma} + U \sum_i n_{i \uparrow} n_{i \downarrow} - \mu \sum_{i \sigma} n_{i \sigma}.
\end{equation}
We work in units where the nearest neighbor hopping $t=1$ and also consider a next-nearest neighbor hopping $t'$.

Observables will be reported as equally-weighted averages over an ensemble of clusters sampling all possible twist conditions; this scheme is known as twist-averaged boundary conditions \cite{kokalj2017,gros1996,lin2001,koretsune2007,qin2016,karakuzu2017,karakuzu2018}. We solve each cluster by a full diagonalization of the Hamiltonian, providing access to all $4^{2\times2} = 256$ energies and eigenstates. All calculations are performed in the grand canonical ensemble; the twist-averaged doping can be tuned continuously even at zero temperature. Our primary object of study will be the twist-averaged density of states
\begin{align}
\rho(\omega) &= \frac{L_x L_y}{(2\pi)^2}\int_0^{\frac{2\pi}{L_x}} \dd{\theta_x} \int_0^{\frac{2\pi}{L_y}} \dd{\theta_y} \rho_{\boldsymbol{\theta}}(\omega) \label{ta_rho}\\
\rho_{\boldsymbol{\theta}}(\omega) &=\begin{multlined}[t]
\frac{1}{Z_{\boldsymbol{\theta}}}\sum_{n m} \abs{\mel**{n}{c_{i\sigma}}{m}}^2 \qty(e^{-\beta E_n} + e^{-\beta E_m}) \\
\times \delta(\omega + E_n - E_m).
\end{multlined}\label{lehmann}
\end{align}
In the last line, $\ket{n}$ and $\ket{m}$ are eigenstates of $H_{\boldsymbol{\theta}}$, with energies $E_n$ and $E_m$, respectively. $Z_{\boldsymbol{\theta}} = \tr e^{-\beta H_{\boldsymbol{\theta}}}$ is the partition function at temperature $T \equiv 1/\beta$. By translational and spin-rotational symmetry, the density of states is the same for every site $i$ and spin $\sigma$.

It is clear from the Lehmann representation, as written in \eqref{lehmann}, that the density of states on an individual cluster is a discrete set of poles. From a calculation of a single $H_{\boldsymbol{\theta}}$, it is therefore difficult to distinguish possible pseudogap effects from the gaps due to the small dimension of the Hilbert space. By contrast, the twist-averaged density of states in \eqref{ta_rho} is supported on a continuous range of frequency and potentially indicates a pseudogap.

\begin{figure}[t]
\includegraphics{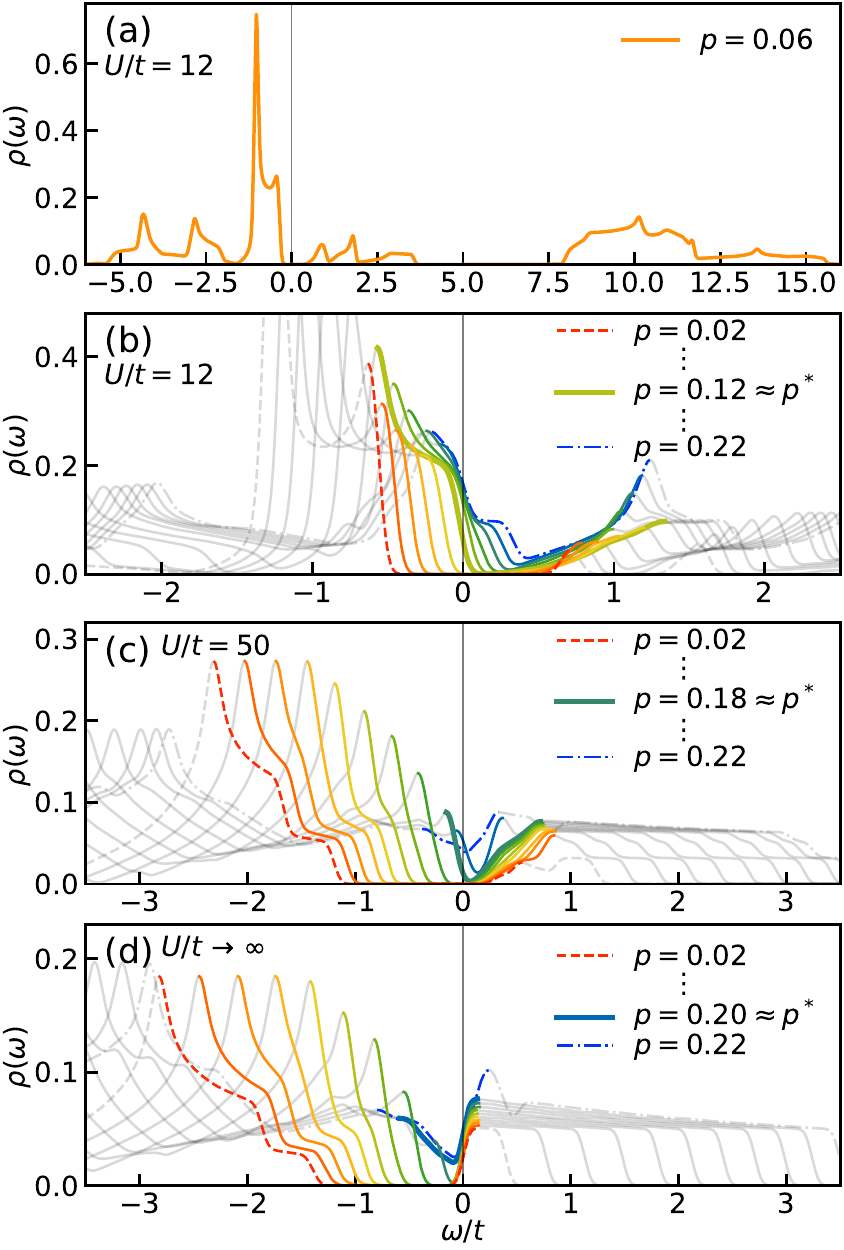}
\caption{Twist-averaged density of states $\rho(\omega)$ at $T=0$. Parameters are $U/t=12$, $t'/t=-0.3$ in (a, b); $U/t=100$, $t'/t=-0.3$ in (c); and $U/t\to\infty$, $t'/t=-0.3$ in (d). The hole doping $p$ in (b-d) varies from $p=0.02$ to $p=0.22$ in increments of $0.02$. For clarity in (b, c, d), the spectra is colored only in the interval between the lowest energy local maxima in the removal and addition spectra. Twist-averaging is performed over $100 \times 100$ values of $\boldsymbol{\theta}$. A very small Gaussian broadening ($\sigma = 0.05$) is applied to make the spectra continuous, and is responsible for the false nonzero $\rho(\omega=0)$ in (d) when $p < p^*$.}
\label{fig:2}
\end{figure}

\emph{Results.---}The density of states for a typical set of parameters $U/t=12$, $t'/t=-0.3$ is plotted in Fig.~\ref{fig:2}(a) for a hole doping $p = 0.06$ and zero temperature. At this interaction strength, the Mott gap is well formed and the upper Hubbard band at energy $\sim U$ is separated distinctly from the lower Hubbard band. The low energy spectrum displays a number of features. Most notably, there is no density of states at zero energy. This is surprising as the system is compressible at this doping, as we have checked by varying the chemical potential. This suppression of density of states at low energy, which we refer to as the pseudogap in our calculations, is not present for all values of doping, temperature, and parameters $U/t$ and $t'/t$. In the following discussion, we map out the region of the temperature-doping phase diagram and parameter space where the pseudogap is present.

In Fig.~\ref{fig:2}(b), we fix the temperature to zero and vary the hole doping. There is a pseudogap for any hole doping $p < p^* \approx 0.12$. As the hole doping is increased from 0 to $p^*$, the leading edge of the removal spectrum moves toward $\omega=0$, approximately linearly with doping. Above the critical doping $p^*$, this edge reaches zero energy and new features in the addition spectrum appear. Similar plots are shown for a very high interaction strength of $U/t = 50$ in Fig.~\ref{fig:2}(c) and for the limit $U/t \to \infty$ in Fig.~\ref{fig:2}(d), indicating that the pseudogap persists in the strongly interacting limit. As the interaction strength increases, the critical doping saturates at a value $p^* \approx 0.2$. Additionally, the pseudogap becomes larger in the removal spectrum and smaller in the addition spectrum. This asymmetry and its parameter dependence implies that multiple energy scales are involved in the pseudogap. Empirically, we find that for large $U/t$, the removal spectrum is gapped up to an energy $\mathcal{O}(t)$ and the addition spectrum is gapped up to an energy $\mathcal{O}\qty(\frac{t^2}{U})$. The persistence of the pseudogap with an energy scale $\mathcal{O}(t)$ in the large $U/t$ limit agrees with previous findings for the Hubbard model from more sophisticated calculations \cite{senechal2004,tremblay2006,kyung2006}.

\begin{figure}[t]
\includegraphics{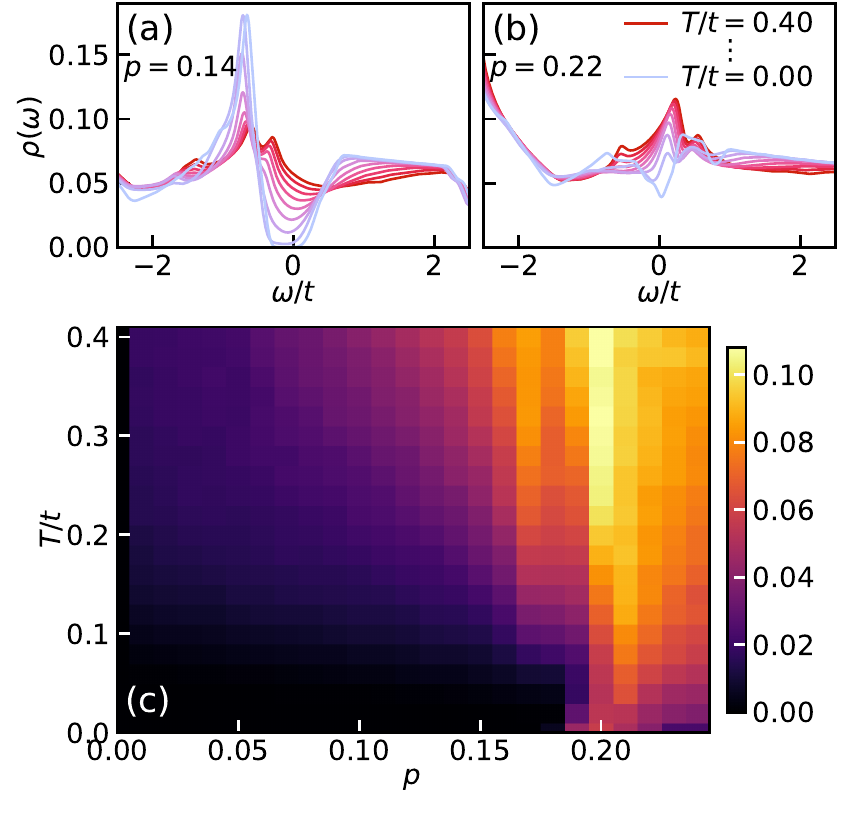}
\caption{(a, b) Temperature dependence of density of states $\rho(\omega)$ for parameters $U/t=50$, $t'/t=-0.3$ and hole doping as indicated. Temperature varies from $T/t = 0.40$ to $0.00$ in steps of $0.05$. (c) Zero energy density of states 
$\rho(\omega=0)$ for a range of hole doping and temperature.}
\label{fig:3}
\end{figure}

We have established that the pseudogap is present below a critical doping $p^*$, with a size that decreases with increasing hole doping. Next, we investigate the temperature dependence. The density of states for hole doping below and above $p^*$ is shown in Fig.~\ref{fig:3}(a) and (b), respectively, for a range of temperatures. Here the interaction strength is $U/t = 50$; we have checked that both smaller and larger interactions give similar behavior. At $p=0.14 < p^*$ in Fig.~\ref{fig:3}(a), the density of states at low energy decreases smoothly as temperature is lowered. By contrast, at $p=0.22 > p^*$ in Fig.~\ref{fig:3}(b), $\rho(\omega)$ is less temperature dependent and the density of states at low energy remains significant down to $T=0$. A 2D plot of $\rho(\omega=0)$ in Fig.~\ref{fig:3}(c) highlights this contrasting behavior over a broad range of hole doping and temperature. For all $p < p^* \approx 0.19$, the pseudogap onsets gradually as temperature is lowered. We cannot pick out a specific onset temperature, but it is evident in Fig.~\ref{fig:3}(c) that any definition of $T^*$ based on $\rho(\omega=0)$ would be a decreasing function of hole doping.

A prominent feature in Fig.~\ref{fig:3}(c) is the nearly temperature independent vertical boundary rising from $T=0$, $p=p^*$. In other words, there is a sharp difference between spectra for doping slightly below and slightly above $p^*$, which persists to high temperatures. Experimentally, such a vertical boundary has been discovered in a recent angle-resolved photoemission (ARPES) study on the bilayer cuprate Bi2212, and found to persist up to at least $310 \mathrm{K}$. (If we estimate that for cuprates, $t \approx 300\mathrm{meV}$, $310 \mathrm{K} \approx 0.09t$.) In this experiment, the spectral function was obtained by Fermi-function-division of the removal spectrum measured by ARPES, rather than by symmetrizing about $\omega=0$ as often performed in previous works. This more accurate analysis showed that as temperature is increased, the pseudogap in the ARPES data does not close at any well-defined temperature and persists to higher temperatures than previously reported. Given that our calculations are not intended to describe cuprates realistically, these similarities between our calculations and recent experiments are striking and suggest the presence of similar strong-correlation effects in both.

\begin{figure}[t]
\includegraphics{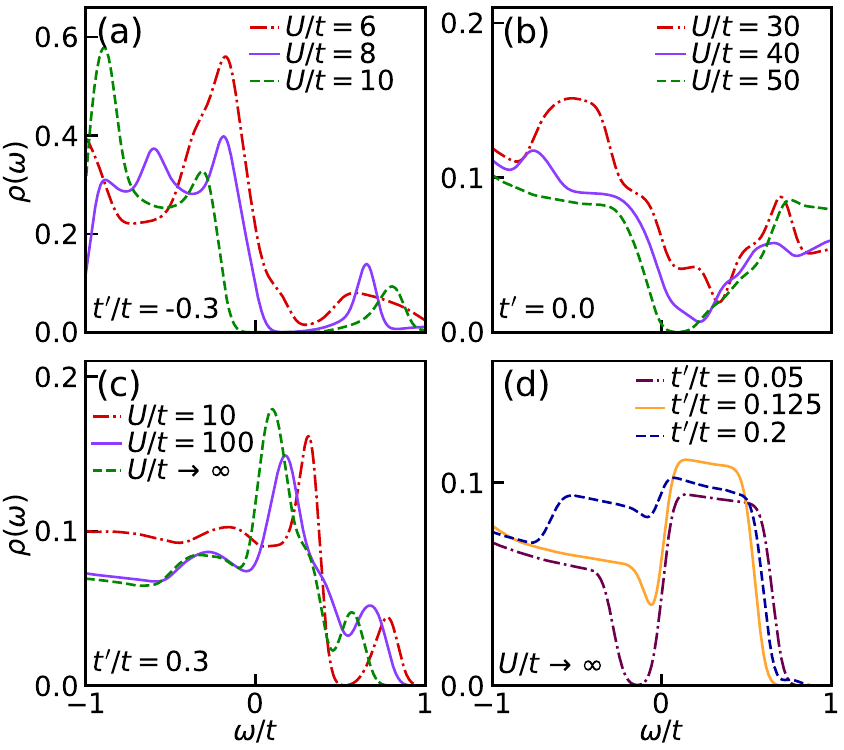}
\caption{Density of states $\rho(\omega)$ at $T=0$ for hole doping $p=0.06$ and parameters as indicated. The critical interaction strength is approximately $U^*/t=8$ in (a) and $U^*/t=40$ in (b). A small Gaussian broadening ($\sigma = 0.05$) is used in all plots and is responsible in (d) for nonzero $\rho(\omega=0)$ when $t'/t=0.05$, since the leading edge in the addition spectra is at $\mathcal{O}\qty(\frac{t^2}{U})$.}
\label{fig:4}
\end{figure}

Finally, we examine the $U/t$ and $t'/t$ dependence of the pseudogap. Our findings can be summarized as follows: when $t'/t < 0.125$, the pseudogap is present if $U > U^*$. The critical interaction strength $U^*$ diverges at $t'/t = 0.125$ and there is no pseudogap at any $U$ when $t'/t \geq 0.125$. Examples are shown in Fig.~\ref{fig:4}. At $t'/t=-0.3$, $U^*/t \approx 8$ (Fig.~\ref{fig:4}(a)); at $t'/t=0$, $U^*/t \approx 40$ (Fig.~\ref{fig:4}(b)); and at $t'/t=0.3$, there is no pseudogap (Fig.~\ref{fig:4}(c)). In Fig.~\ref{fig:4}(d) we see that for $U/t \to \infty$, the zero energy density of states is suppressed only when $t'/t < 0.125$.

 In the Hubbard model, a particle-hole transformation $c_i^\dagger \to (-1)^{x_i + y_i} c_i$ inverts the sign of $t'/t$. Therefore, Fig.~\ref{fig:4}(c) can be understood as the density of states, with energy axis flipped, for $t'/t = -0.3$ and electron-doping. This particle-hole transformation shows that when $t'/t < -0.125$ the pseudogap is absent for electron-doping, regardless of the interaction strength. The fact electron-doping with negative $t'/t$ (or hole-doping with positive $t'/t$) is unfavorable for the pseudogap agrees with existing literature \cite{senechal2004,kyung2006,macridin2006,liebsch2009,kohno2014,wu2018}, although we note that in many of these studies a pseudogap is found for $t'/t = 0$ without requiring as large of a interaction strength as in Fig.~\ref{fig:4}(b). While there are certainly shortcomings of calculations for a $2\times2$ cluster, its advantage lies in being simple enough that the mechanism of the pseudogap can be determined unambiguously, as we now discuss.

\emph{Mechanism and Discussion.---}We have demonstrated that the density of states for a $2\times2$ cluster with twist-averaged boundary conditions exhibits a pseudogap with characteristics highly reminiscent of findings in more elaborate calculations of the Hubbard model. We now turn to the question of the mechanism of the pseudogap. Given that all many-body energies and eigenstates are calculated exactly for every $H_{\boldsymbol{\theta}}$, there is no technical hindrance to determining the answer.

Our starting point is to consider the Lehmann representation of the density of states in \eqref{lehmann}. At zero temperature,
\begin{align}
\rho_{\boldsymbol{\theta}}(\omega) &= \rho_{\boldsymbol{\theta}}^<(\omega) + \rho_{\boldsymbol{\theta}}^>(\omega) \\
\rho_{\boldsymbol{\theta}}^<(\omega) &= \sum_{n} \abs{\mel**{n}{c_{i\sigma}}{G}}^2 \delta(\omega + (E_n - E_G))) \label{lehmann_rem}\\
\rho_{\boldsymbol{\theta}}^>(\omega) &= \sum_{n} \abs{\mel**{n}{c_{i\sigma}^\dagger}{G}}^2 \delta(\omega - (E_n - E_G)), \label{lehmann_add}
\end{align}
where $\ket{G}$ is the ground state of $H_{\boldsymbol{\theta}}$. (The generalization of this discussion to degenerate ground states is trivial.) Here, $\rho_{\boldsymbol{\theta}}^<$ and $\rho_{\boldsymbol{\theta}}^>$ refer to the removal and addition spectra, respectively. From the spectral representation, it is clear that two conditions are necessary to have any low energy spectral weight in the removal (addition) spectrum. First, there must be low energy excited states with one less (more) electron than in $\ket{G}$. Second, these excited states must have nonzero overlap with the ground state with one electron removed (added). A full suppression of low energy spectral weight, in Fig.~\ref{fig:2} for instance, requires one of these conditions to be violated.

\begin{figure}[t]
\includegraphics{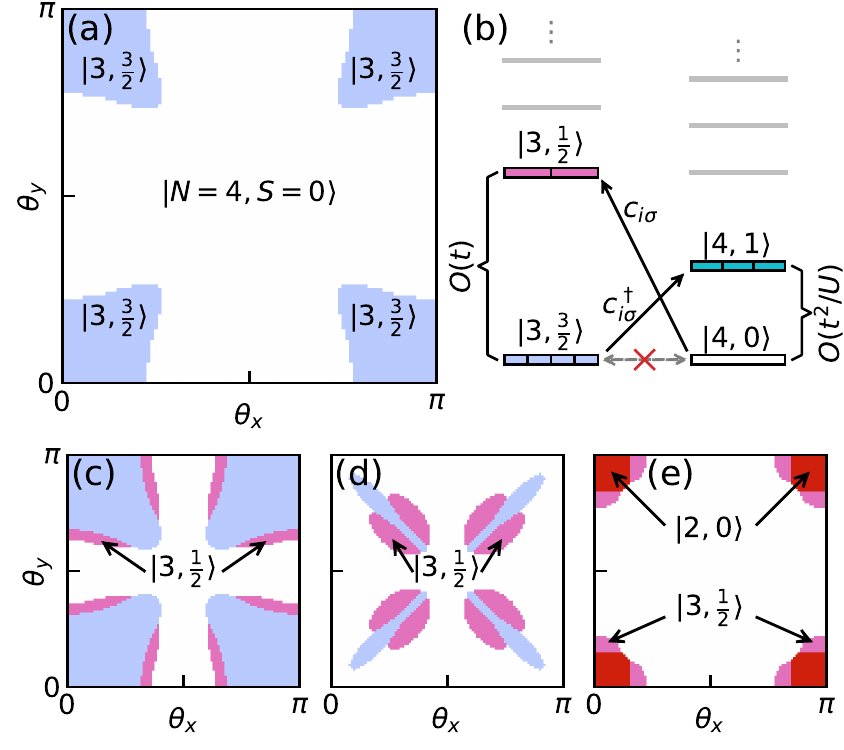}
\caption{(a) Ground state total charge $N$ and total spin $S$ for $H_{\boldsymbol{\theta}}$, plotted over a $100\times100$ grid of $\boldsymbol{\theta}$. The pseudogap is present here for the parameters $U/t=12$, $t'/t=-0.3$, $p=0.06$, and $T=0$. (b) Schematic of the many-body excitation spectrum. (c-e) Like (a), but with one parameter modified in each: (c) $p=0.15 > p^*$, (d) $U/t=7<U^*$, and (e) $t'/t=0.3$. There is no pseudogap present in (c-e). Unlabelled light blue and white regions indicate $N=3$, $S=\frac{3}{2}$ and $N=4$, $S=0$, as in (a).}
\label{fig:5}
\end{figure}

Because we are dealing with the twist-averaged density of states $\rho(\omega)$, these conditions should be checked for all ${\boldsymbol{\theta}}$. In Fig.~\ref{fig:5}(a), we show the total density and total spin quantum numbers of the ground state of each $H_{\boldsymbol{\theta}}$. Because eigenstates and energies evolve continuously with ${\boldsymbol{\theta}}$, the first condition is satisfied when ${\boldsymbol{\theta}}$ is close to a border where the total density changes by 1. For example, if ${\boldsymbol{\theta}}$ is near such a border on the side where the ground state has $N=4$, there must be a low energy excited state with $N=3$. If there is any low energy density of states in $\rho(\omega)$, it must be due to the contributions of $\rho_{\boldsymbol{\theta}}(\omega)$ with ${\boldsymbol{\theta}}$ near these boundaries.

However, we can infer from the same figure that while configurations near boundaries satisfy the first condition, the second condition is violated. This is because the $N=3$ state has total spin $S=\frac{3}{2}$ but the $N=4$ state has total spin $S=0$. The matrix elements in \eqref{lehmann_rem} and \eqref{lehmann_add} must vanish because of spin selection rules; the $N=3$ state is orthogonal to the $N=4$ state with one electron removed.

This orthogonality is the cause of the pseudogap in our calculations of the $2\times2$ cluster. To see this, we sketch in Fig.~\ref{fig:5}(b) the low energy many-body spectrum of $H_{\boldsymbol{\theta}}$ for ${\boldsymbol{\theta}}$ near a border in Fig.~\ref{fig:5}(a). Tuning ${\boldsymbol{\theta}}$ across the border tips the balance between the $\ket{3,\frac{3}{2}}$ and $\ket{4,0}$ states. For the removal spectrum, consider ${\boldsymbol{\theta}}$ where the ground state is $\ket{4,0}$. Because $\mel{3,\frac{3}{2}}{c_{i\sigma}}{4,0} = 0$, the lowest energy state appearing in $\rho_{\boldsymbol{\theta}}^<(\omega)$ is $\ket{3,\frac{1}{2}}$. This state has energy $\mathcal{O}(t)$ higher than $\ket{3,\frac{3}{2}}$ as we will discuss shortly. As $\ket{4,0}$ is the ground state, $E_{3,\frac{1}{2}} > E_{3,\frac{3}{2}} > E_{4,0}$. Therefore, even after integrating over ${\boldsymbol{\theta}}$, the removal spectrum must be gapped by $\mathcal{O}(t)$. An identical argument applies for the addition spectra, where the lowest energy visible state is $\ket{4,1}$, which is $\mathcal{O}\qty(\frac{t^2}{U})$ higher in energy than $\ket{4,0}$. This description, of a $\mathcal{O}(t)$ gap in the removal spectrum and $\mathcal{O}\qty(\frac{t^2}{U})$ in the addition spectrum, is precisely what we observed previously in Fig.~\ref{fig:2}.

It is straightforward to understand the $\mathcal{O}\qty(\frac{t^2}{U})$ energy difference between the singlet $\ket{4,0}$ and triplet $\ket{4,1}$ states. For $N=4$, at large $U/t$, kinetic motion is frozen and the low-energy effective Hamiltonian only involves $\mathcal{O}\qty(\frac{t^2}{U})$ exchange interactions. The $\mathcal{O}(t)$ difference between $\ket{3,\frac{3}{2}}$ and $\ket{3,\frac{1}{2}}$ is less simple. It is helpful to view the $2\times2$ cluster as an isolated plaquette with hoppings $t_x = 2 t \cos \theta_x$, $t_y = 2 t \cos \theta_y$, $t_{diag} = 4 t' \cos \theta_x \cos \theta_y$, as determined by adding together equivalent bonds in Fig.~\ref{fig:1}. We consider small doping, such that the border in Fig.~\ref{fig:5}(a) is near $\boldsymbol{\theta}=(0,0)$, so $t_x \approx t_y \approx 2t$ and $t_{diag} \approx 4 t'$.
In the limit $U/t \to \infty$, the fact that the fully polarized $\ket{3,\frac{3}{2}}$ state is lower in energy than $\ket{3,\frac{1}{2}}$ may be understood as an instance of Nagaoka ferromagnetism \cite{nagaoka}. Critically, Nagaoka's arguments require the product of negative hoppings $-t_{i j}$ around any loop to be positive; this condition is fulfilled if $t_{diag} \leq 0$.

Indeed, we find in Fig.~\ref{fig:5}(e) that the fully polarized state is absent for positive $t_{diag}/t_x \approx 2 t'/t$. Instead, there are low energy $\ket{3,\frac{1}{2}}$ states, for which the matrix element $\mel{3,\frac{1}{2}}{c_{i\sigma}}{4,0}$ is nonzero, giving rise to density of states at low energy. The critical value of $t'/t$ where the $N=3$ ground state changes total spin can be determined analytically for a $2\times2$ plaquette in the limit $U/t \to \infty$ \cite{buterakos2019}. The requirement is $t_{diag}<\frac{t_x t_y}{3 t_x + t_y}$ (where $t_x > t_y$), which translates to $t'/t < 0.125$ at small doping. This is precisely what we determined previously in Fig.~\ref{fig:4}(d).

In the 4-site plaquette, the spin-$\frac{3}{2}$ state remains the ground state for finite values of $U/t$ above a critical $U^*/t$. For $t_x=t_y$ and $t_{diag}=0$, the critical value is $U^*/t_x \approx 18.7$, in close agreement with our estimate of $U^*/(2 t) \approx 20$ in Fig.~\ref{fig:4}(b). The spin-$\frac{3}{2}$ state is stabilized further by negative values of $t_{diag}/t_x$; hence we see in Fig.~\ref{fig:5}(a) that the ground state can be spin-$\frac{3}{2}$ even for $U/t=12$. Reducing to $U/t=7$ in Fig.~\ref{fig:5}(d) leads to regions of $\boldsymbol{\theta}$ where the ground state is spin-$\frac{1}{2}$. Similarly in Fig.~\ref{fig:5}(c), we see that increasing doping shifts the borders in $\boldsymbol{\theta}$ that are important to analyzing the low energy density of states. This gives rise to spin-$\frac{1}{2}$ regions because when $\boldsymbol{\theta}$ is near $(\frac{\pi}{2},0)$ or $(0,\frac{\pi}{2})$, the 4-site plaquette becomes highly anisotropic, which is known to be energetically unfavorable for the spin-$\frac{3}{2}$ state \cite{buterakos2019}. For the parameters $U/t=12$ $t'/t=-0.3$, the first spin-$\frac{1}{2}$ regions appear at $p^*=0.1225(2)$ hole doping, in exact agreement with Fig.~\ref{fig:2}(b).

By analyzing the ground state quantum numbers and many-body excitation spectrum as a function of twist parameter $\boldsymbol{\theta}$, we have understood the parameter dependence and the energetics of the pseudogap. From the excitation spectrum sketched in Fig.~\ref{fig:5}(b), we can also infer that the pseudogap fills in when $T \sim \frac{t^2}{U}$. As temperature rises, the thermal mixed state begins to include triplet $\ket{4,1}$ states. Transitions to and from $\ket{3,\frac{3}{2}}$ are no longer forbidden by selection rules and lead to low energy spectral weight. Such an association between the onset of the pseudogap with the shift of probability between triplet and singlet states has been reported previously in cluster DMFT calculations \cite{gull2008,sordi2012srep}. Our results and analysis here of the low energy spin-$\frac{3}{2}$ states provide a clear explanation.

The important open question that remains is whether our findings generalize to the Hubbard model in the thermodynamic limit. The $2\times2$ cluster certainly is missing many important qualitative phenomena related to longer-distance correlations. However, the particular parameter, doping, and temperature dependences we have observed and their consistency with more sophisticated calculations suggests that the aspects of pseudogap physics captured by the small cluster are present also in larger systems. The key point is the orthogonality, by spin selection rules, between bare electrons and spin-$\frac{3}{2}$ excitations. For a 4-site system, the low energy of spin-$\frac{3}{2}$ states may be understood as an application of Nagaoka's theorem. It is highly unlikely that Nagaoka physics is relevant to the Hubbard model on large systems in the parameter regime $U/t \sim 8$ most commonly studied. However, this does not preclude a more general reason for the presence of low-energy spin-$\frac{3}{2}$ excitations, which, if lower in energy than a substantial portion of spin-$\frac{1}{2}$ excitations, would give rise to a pseudogap by the same mechanism that we have found here. In fact, there is evidence from numerically exact quantum Monte Carlo calculations on $8\times8$ clusters showing that spectroscopically invisible spin-$\frac{3}{2}$ excitations are responsible for missing features expected in the single-particle spectral function \cite{grober2000}. Similar calculations at lower temperatures on large systems, with measurements of the Green's function of composite spin-$\frac{3}{2}$ operators, could clarify the relevance of the small cluster physics revealed here to the pseudogap of the Hubbard model in the thermodynamic limit.

\begin{acknowledgments}
EWH thanks Philip Phillips, Bryan Clark, Thomas Devereaux, Yao Wang, Yu He, and Sudi Chen for helpful discussions and was supported by the Gordon and Betty Moore Foundation EPiQS Initiative through the grants GBMF 4305 and GBMF 8691.
\end{acknowledgments}

\bibliographystyle{apsrev4-2}

\begin{thebibliography}{48}%
\makeatletter
\providecommand \@ifxundefined [1]{%
 \@ifx{#1\undefined}
}%
\providecommand \@ifnum [1]{%
 \ifnum #1\expandafter \@firstoftwo
 \else \expandafter \@secondoftwo
 \fi
}%
\providecommand \@ifx [1]{%
 \ifx #1\expandafter \@firstoftwo
 \else \expandafter \@secondoftwo
 \fi
}%
\providecommand \natexlab [1]{#1}%
\providecommand \enquote  [1]{``#1''}%
\providecommand \bibnamefont  [1]{#1}%
\providecommand \bibfnamefont [1]{#1}%
\providecommand \citenamefont [1]{#1}%
\providecommand \href@noop [0]{\@secondoftwo}%
\providecommand \href [0]{\begingroup \@sanitize@url \@href}%
\providecommand \@href[1]{\@@startlink{#1}\@@href}%
\providecommand \@@href[1]{\endgroup#1\@@endlink}%
\providecommand \@sanitize@url [0]{\catcode `\\12\catcode `\$12\catcode
  `\&12\catcode `\#12\catcode `\^12\catcode `\_12\catcode `\%12\relax}%
\providecommand \@@startlink[1]{}%
\providecommand \@@endlink[0]{}%
\providecommand \url  [0]{\begingroup\@sanitize@url \@url }%
\providecommand \@url [1]{\endgroup\@href {#1}{\urlprefix }}%
\providecommand \urlprefix  [0]{URL }%
\providecommand \Eprint [0]{\href }%
\providecommand \doibase [0]{https://doi.org/}%
\providecommand \selectlanguage [0]{\@gobble}%
\providecommand \bibinfo  [0]{\@secondoftwo}%
\providecommand \bibfield  [0]{\@secondoftwo}%
\providecommand \translation [1]{[#1]}%
\providecommand \BibitemOpen [0]{}%
\providecommand \bibitemStop [0]{}%
\providecommand \bibitemNoStop [0]{.\EOS\space}%
\providecommand \EOS [0]{\spacefactor3000\relax}%
\providecommand \BibitemShut  [1]{\csname bibitem#1\endcsname}%
\let\auto@bib@innerbib\@empty
\bibitem [{\citenamefont {Keimer}\ \emph {et~al.}(2015)\citenamefont {Keimer},
  \citenamefont {Kivelson}, \citenamefont {Norman}, \citenamefont {Uchida},\
  and\ \citenamefont {Zaanen}}]{keimer2015}%
  \BibitemOpen
  \bibfield  {author} {\bibinfo {author} {\bibfnamefont {B.}~\bibnamefont
  {Keimer}}, \bibinfo {author} {\bibfnamefont {S.~A.}\ \bibnamefont
  {Kivelson}}, \bibinfo {author} {\bibfnamefont {M.~R.}\ \bibnamefont
  {Norman}}, \bibinfo {author} {\bibfnamefont {S.}~\bibnamefont {Uchida}},\
  and\ \bibinfo {author} {\bibfnamefont {J.}~\bibnamefont {Zaanen}},\ }\href
  {https://doi.org/10.1038/nature14165} {\bibfield  {journal} {\bibinfo
  {journal} {Nature}\ }\textbf {\bibinfo {volume} {518}},\ \bibinfo {pages}
  {179} (\bibinfo {year} {2015})}\BibitemShut {NoStop}%
\bibitem [{\citenamefont {Fradkin}\ \emph {et~al.}(2015)\citenamefont
  {Fradkin}, \citenamefont {Kivelson},\ and\ \citenamefont
  {Tranquada}}]{fradkin2015}%
  \BibitemOpen
  \bibfield  {author} {\bibinfo {author} {\bibfnamefont {E.}~\bibnamefont
  {Fradkin}}, \bibinfo {author} {\bibfnamefont {S.~A.}\ \bibnamefont
  {Kivelson}},\ and\ \bibinfo {author} {\bibfnamefont {J.~M.}\ \bibnamefont
  {Tranquada}},\ }\href {https://doi.org/10.1103/RevModPhys.87.457} {\bibfield
  {journal} {\bibinfo  {journal} {Rev. Mod. Phys.}\ }\textbf {\bibinfo {volume}
  {87}},\ \bibinfo {pages} {457} (\bibinfo {year} {2015})}\BibitemShut
  {NoStop}%
\bibitem [{\citenamefont {Prelov\ifmmode~\check{s}\else \v{s}\fi{}ek}\ \emph
  {et~al.}(1999)\citenamefont {Prelov\ifmmode~\check{s}\else \v{s}\fi{}ek},
  \citenamefont {Jakli\ifmmode~\check{c}\else \v{c}\fi{}},\ and\ \citenamefont
  {Bedell}}]{prelovsek1999}%
  \BibitemOpen
  \bibfield  {author} {\bibinfo {author} {\bibfnamefont {P.}~\bibnamefont
  {Prelov\ifmmode~\check{s}\else \v{s}\fi{}ek}}, \bibinfo {author}
  {\bibfnamefont {J.}~\bibnamefont {Jakli\ifmmode~\check{c}\else \v{c}\fi{}}},\
  and\ \bibinfo {author} {\bibfnamefont {K.}~\bibnamefont {Bedell}},\ }\href
  {https://doi.org/10.1103/PhysRevB.60.40} {\bibfield  {journal} {\bibinfo
  {journal} {Phys. Rev. B}\ }\textbf {\bibinfo {volume} {60}},\ \bibinfo
  {pages} {40} (\bibinfo {year} {1999})}\BibitemShut {NoStop}%
\bibitem [{\citenamefont {Maier}\ \emph {et~al.}(2000)\citenamefont {Maier},
  \citenamefont {Jarrell}, \citenamefont {Pruschke},\ and\ \citenamefont
  {Keller}}]{maier2000}%
  \BibitemOpen
  \bibfield  {author} {\bibinfo {author} {\bibfnamefont {T.}~\bibnamefont
  {Maier}}, \bibinfo {author} {\bibfnamefont {M.}~\bibnamefont {Jarrell}},
  \bibinfo {author} {\bibfnamefont {T.}~\bibnamefont {Pruschke}},\ and\
  \bibinfo {author} {\bibfnamefont {J.}~\bibnamefont {Keller}},\ }\href
  {https://doi.org/10.1007/s100510050077} {\bibfield  {journal} {\bibinfo
  {journal} {Eur. Phys. J. B}\ }\textbf {\bibinfo {volume} {13}},\ \bibinfo
  {pages} {613} (\bibinfo {year} {2000})}\BibitemShut {NoStop}%
\bibitem [{\citenamefont {Huscroft}\ \emph {et~al.}(2001)\citenamefont
  {Huscroft}, \citenamefont {Jarrell}, \citenamefont {Maier}, \citenamefont
  {Moukouri},\ and\ \citenamefont {Tahvildarzadeh}}]{maier2001}%
  \BibitemOpen
  \bibfield  {author} {\bibinfo {author} {\bibfnamefont {C.}~\bibnamefont
  {Huscroft}}, \bibinfo {author} {\bibfnamefont {M.}~\bibnamefont {Jarrell}},
  \bibinfo {author} {\bibfnamefont {T.}~\bibnamefont {Maier}}, \bibinfo
  {author} {\bibfnamefont {S.}~\bibnamefont {Moukouri}},\ and\ \bibinfo
  {author} {\bibfnamefont {A.~N.}\ \bibnamefont {Tahvildarzadeh}},\ }\href
  {https://doi.org/10.1103/PhysRevLett.86.139} {\bibfield  {journal} {\bibinfo
  {journal} {Phys. Rev. Lett.}\ }\textbf {\bibinfo {volume} {86}},\ \bibinfo
  {pages} {139} (\bibinfo {year} {2001})}\BibitemShut {NoStop}%
\bibitem [{\citenamefont {Jarrell}\ \emph {et~al.}(2001)\citenamefont
  {Jarrell}, \citenamefont {Maier}, \citenamefont {Hettler},\ and\
  \citenamefont {Tahvildarzadeh}}]{jarrell2001}%
  \BibitemOpen
  \bibfield  {author} {\bibinfo {author} {\bibfnamefont {M.}~\bibnamefont
  {Jarrell}}, \bibinfo {author} {\bibfnamefont {T.}~\bibnamefont {Maier}},
  \bibinfo {author} {\bibfnamefont {M.~H.}\ \bibnamefont {Hettler}},\ and\
  \bibinfo {author} {\bibfnamefont {A.~N.}\ \bibnamefont {Tahvildarzadeh}},\
  }\href {https://doi.org/10.1209/epl/i2001-00557-x} {\bibfield  {journal}
  {\bibinfo  {journal} {Europhys. Lett.}\ }\textbf {\bibinfo {volume} {56}},\
  \bibinfo {pages} {563} (\bibinfo {year} {2001})}\BibitemShut {NoStop}%
\bibitem [{\citenamefont {Haule}\ \emph {et~al.}(2002)\citenamefont {Haule},
  \citenamefont {Rosch}, \citenamefont {Kroha},\ and\ \citenamefont
  {W\"olfle}}]{haule2002}%
  \BibitemOpen
  \bibfield  {author} {\bibinfo {author} {\bibfnamefont {K.}~\bibnamefont
  {Haule}}, \bibinfo {author} {\bibfnamefont {A.}~\bibnamefont {Rosch}},
  \bibinfo {author} {\bibfnamefont {J.}~\bibnamefont {Kroha}},\ and\ \bibinfo
  {author} {\bibfnamefont {P.}~\bibnamefont {W\"olfle}},\ }\href
  {https://doi.org/10.1103/PhysRevLett.89.236402} {\bibfield  {journal}
  {\bibinfo  {journal} {Phys. Rev. Lett.}\ }\textbf {\bibinfo {volume} {89}},\
  \bibinfo {pages} {236402} (\bibinfo {year} {2002})}\BibitemShut {NoStop}%
\bibitem [{\citenamefont {Haule}\ \emph {et~al.}(2003)\citenamefont {Haule},
  \citenamefont {Rosch}, \citenamefont {Kroha},\ and\ \citenamefont
  {W\"olfle}}]{haule2003}%
  \BibitemOpen
  \bibfield  {author} {\bibinfo {author} {\bibfnamefont {K.}~\bibnamefont
  {Haule}}, \bibinfo {author} {\bibfnamefont {A.}~\bibnamefont {Rosch}},
  \bibinfo {author} {\bibfnamefont {J.}~\bibnamefont {Kroha}},\ and\ \bibinfo
  {author} {\bibfnamefont {P.}~\bibnamefont {W\"olfle}},\ }\href
  {https://doi.org/10.1103/PhysRevB.68.155119} {\bibfield  {journal} {\bibinfo
  {journal} {Phys. Rev. B}\ }\textbf {\bibinfo {volume} {68}},\ \bibinfo
  {pages} {155119} (\bibinfo {year} {2003})}\BibitemShut {NoStop}%
\bibitem [{\citenamefont {Stanescu}\ and\ \citenamefont
  {Phillips}(2003)}]{stanescu2003}%
  \BibitemOpen
  \bibfield  {author} {\bibinfo {author} {\bibfnamefont {T.~D.}\ \bibnamefont
  {Stanescu}}\ and\ \bibinfo {author} {\bibfnamefont {P.}~\bibnamefont
  {Phillips}},\ }\href {https://doi.org/10.1103/PhysRevLett.91.017002}
  {\bibfield  {journal} {\bibinfo  {journal} {Phys. Rev. Lett.}\ }\textbf
  {\bibinfo {volume} {91}},\ \bibinfo {pages} {017002} (\bibinfo {year}
  {2003})}\BibitemShut {NoStop}%
\bibitem [{\citenamefont {S\'en\'echal}\ and\ \citenamefont
  {Tremblay}(2004)}]{senechal2004}%
  \BibitemOpen
  \bibfield  {author} {\bibinfo {author} {\bibfnamefont {D.}~\bibnamefont
  {S\'en\'echal}}\ and\ \bibinfo {author} {\bibfnamefont {A.-M.~S.}\
  \bibnamefont {Tremblay}},\ }\href
  {https://doi.org/10.1103/PhysRevLett.92.126401} {\bibfield  {journal}
  {\bibinfo  {journal} {Phys. Rev. Lett.}\ }\textbf {\bibinfo {volume} {92}},\
  \bibinfo {pages} {126401} (\bibinfo {year} {2004})}\BibitemShut {NoStop}%
\bibitem [{\citenamefont {Sadovskii}\ \emph {et~al.}(2005)\citenamefont
  {Sadovskii}, \citenamefont {Nekrasov}, \citenamefont {Kuchinskii},
  \citenamefont {Pruschke},\ and\ \citenamefont {Anisimov}}]{sadovskii2005}%
  \BibitemOpen
  \bibfield  {author} {\bibinfo {author} {\bibfnamefont {M.~V.}\ \bibnamefont
  {Sadovskii}}, \bibinfo {author} {\bibfnamefont {I.~A.}\ \bibnamefont
  {Nekrasov}}, \bibinfo {author} {\bibfnamefont {E.~Z.}\ \bibnamefont
  {Kuchinskii}}, \bibinfo {author} {\bibfnamefont {T.}~\bibnamefont
  {Pruschke}},\ and\ \bibinfo {author} {\bibfnamefont {V.~I.}\ \bibnamefont
  {Anisimov}},\ }\href {https://doi.org/10.1103/PhysRevB.72.155105} {\bibfield
  {journal} {\bibinfo  {journal} {Phys. Rev. B}\ }\textbf {\bibinfo {volume}
  {72}},\ \bibinfo {pages} {155105} (\bibinfo {year} {2005})}\BibitemShut
  {NoStop}%
\bibitem [{\citenamefont {Stanescu}\ and\ \citenamefont
  {Kotliar}(2006)}]{stanescu2006}%
  \BibitemOpen
  \bibfield  {author} {\bibinfo {author} {\bibfnamefont {T.~D.}\ \bibnamefont
  {Stanescu}}\ and\ \bibinfo {author} {\bibfnamefont {G.}~\bibnamefont
  {Kotliar}},\ }\href {https://doi.org/10.1103/PhysRevB.74.125110} {\bibfield
  {journal} {\bibinfo  {journal} {Phys. Rev. B}\ }\textbf {\bibinfo {volume}
  {74}},\ \bibinfo {pages} {125110} (\bibinfo {year} {2006})}\BibitemShut
  {NoStop}%
\bibitem [{\citenamefont {Tremblay}\ \emph {et~al.}(2006)\citenamefont
  {Tremblay}, \citenamefont {Kyung},\ and\ \citenamefont
  {Sénéchal}}]{tremblay2006}%
  \BibitemOpen
  \bibfield  {author} {\bibinfo {author} {\bibfnamefont {A.-M.~S.}\
  \bibnamefont {Tremblay}}, \bibinfo {author} {\bibfnamefont {B.}~\bibnamefont
  {Kyung}},\ and\ \bibinfo {author} {\bibfnamefont {D.}~\bibnamefont
  {Sénéchal}},\ }\href {https://doi.org/10.1063/1.2199446} {\bibfield
  {journal} {\bibinfo  {journal} {Low Temp. Phys.}\ }\textbf {\bibinfo {volume}
  {32}},\ \bibinfo {pages} {424} (\bibinfo {year} {2006})}\BibitemShut
  {NoStop}%
\bibitem [{\citenamefont {Kyung}\ \emph {et~al.}(2006)\citenamefont {Kyung},
  \citenamefont {Kancharla}, \citenamefont {S\'en\'echal}, \citenamefont
  {Tremblay}, \citenamefont {Civelli},\ and\ \citenamefont
  {Kotliar}}]{kyung2006}%
  \BibitemOpen
  \bibfield  {author} {\bibinfo {author} {\bibfnamefont {B.}~\bibnamefont
  {Kyung}}, \bibinfo {author} {\bibfnamefont {S.~S.}\ \bibnamefont
  {Kancharla}}, \bibinfo {author} {\bibfnamefont {D.}~\bibnamefont
  {S\'en\'echal}}, \bibinfo {author} {\bibfnamefont {A.-M.~S.}\ \bibnamefont
  {Tremblay}}, \bibinfo {author} {\bibfnamefont {M.}~\bibnamefont {Civelli}},\
  and\ \bibinfo {author} {\bibfnamefont {G.}~\bibnamefont {Kotliar}},\ }\href
  {https://doi.org/10.1103/PhysRevB.73.165114} {\bibfield  {journal} {\bibinfo
  {journal} {Phys. Rev. B}\ }\textbf {\bibinfo {volume} {73}},\ \bibinfo
  {pages} {165114} (\bibinfo {year} {2006})}\BibitemShut {NoStop}%
\bibitem [{\citenamefont {Macridin}\ \emph {et~al.}(2006)\citenamefont
  {Macridin}, \citenamefont {Jarrell}, \citenamefont {Maier}, \citenamefont
  {Kent},\ and\ \citenamefont {D'Azevedo}}]{macridin2006}%
  \BibitemOpen
  \bibfield  {author} {\bibinfo {author} {\bibfnamefont {A.}~\bibnamefont
  {Macridin}}, \bibinfo {author} {\bibfnamefont {M.}~\bibnamefont {Jarrell}},
  \bibinfo {author} {\bibfnamefont {T.}~\bibnamefont {Maier}}, \bibinfo
  {author} {\bibfnamefont {P.~R.~C.}\ \bibnamefont {Kent}},\ and\ \bibinfo
  {author} {\bibfnamefont {E.}~\bibnamefont {D'Azevedo}},\ }\href
  {https://doi.org/10.1103/PhysRevLett.97.036401} {\bibfield  {journal}
  {\bibinfo  {journal} {Phys. Rev. Lett.}\ }\textbf {\bibinfo {volume} {97}},\
  \bibinfo {pages} {036401} (\bibinfo {year} {2006})}\BibitemShut {NoStop}%
\bibitem [{\citenamefont {Gull}\ \emph {et~al.}(2008)\citenamefont {Gull},
  \citenamefont {Werner}, \citenamefont {Wang}, \citenamefont {Troyer},\ and\
  \citenamefont {Millis}}]{gull2008}%
  \BibitemOpen
  \bibfield  {author} {\bibinfo {author} {\bibfnamefont {E.}~\bibnamefont
  {Gull}}, \bibinfo {author} {\bibfnamefont {P.}~\bibnamefont {Werner}},
  \bibinfo {author} {\bibfnamefont {X.}~\bibnamefont {Wang}}, \bibinfo {author}
  {\bibfnamefont {M.}~\bibnamefont {Troyer}},\ and\ \bibinfo {author}
  {\bibfnamefont {A.~J.}\ \bibnamefont {Millis}},\ }\href
  {https://doi.org/10.1209/0295-5075/84/37009} {\bibfield  {journal} {\bibinfo
  {journal} {Europhys. Lett.}\ }\textbf {\bibinfo {volume} {84}},\ \bibinfo
  {pages} {37009} (\bibinfo {year} {2008})}\BibitemShut {NoStop}%
\bibitem [{\citenamefont {Liebsch}\ and\ \citenamefont
  {Tong}(2009)}]{liebsch2009}%
  \BibitemOpen
  \bibfield  {author} {\bibinfo {author} {\bibfnamefont {A.}~\bibnamefont
  {Liebsch}}\ and\ \bibinfo {author} {\bibfnamefont {N.-H.}\ \bibnamefont
  {Tong}},\ }\href {https://doi.org/10.1103/PhysRevB.80.165126} {\bibfield
  {journal} {\bibinfo  {journal} {Phys. Rev. B}\ }\textbf {\bibinfo {volume}
  {80}},\ \bibinfo {pages} {165126} (\bibinfo {year} {2009})}\BibitemShut
  {NoStop}%
\bibitem [{\citenamefont {Vidhyadhiraja}\ \emph {et~al.}(2009)\citenamefont
  {Vidhyadhiraja}, \citenamefont {Macridin}, \citenamefont
  {\ifmmode~\mbox{\c{S}}\else \c{S}\fi{}en}, \citenamefont {Jarrell},\ and\
  \citenamefont {Ma}}]{vidhyadhiraja2009}%
  \BibitemOpen
  \bibfield  {author} {\bibinfo {author} {\bibfnamefont {N.~S.}\ \bibnamefont
  {Vidhyadhiraja}}, \bibinfo {author} {\bibfnamefont {A.}~\bibnamefont
  {Macridin}}, \bibinfo {author} {\bibfnamefont {C.}~\bibnamefont
  {\ifmmode~\mbox{\c{S}}\else \c{S}\fi{}en}}, \bibinfo {author} {\bibfnamefont
  {M.}~\bibnamefont {Jarrell}},\ and\ \bibinfo {author} {\bibfnamefont
  {M.}~\bibnamefont {Ma}},\ }\href
  {https://doi.org/10.1103/PhysRevLett.102.206407} {\bibfield  {journal}
  {\bibinfo  {journal} {Phys. Rev. Lett.}\ }\textbf {\bibinfo {volume} {102}},\
  \bibinfo {pages} {206407} (\bibinfo {year} {2009})}\BibitemShut {NoStop}%
\bibitem [{\citenamefont {Ferrero}\ \emph
  {et~al.}(2009{\natexlab{a}})\citenamefont {Ferrero}, \citenamefont
  {Cornaglia}, \citenamefont {De~Leo}, \citenamefont {Parcollet}, \citenamefont
  {Kotliar},\ and\ \citenamefont {Georges}}]{ferrero2009}%
  \BibitemOpen
  \bibfield  {author} {\bibinfo {author} {\bibfnamefont {M.}~\bibnamefont
  {Ferrero}}, \bibinfo {author} {\bibfnamefont {P.~S.}\ \bibnamefont
  {Cornaglia}}, \bibinfo {author} {\bibfnamefont {L.}~\bibnamefont {De~Leo}},
  \bibinfo {author} {\bibfnamefont {O.}~\bibnamefont {Parcollet}}, \bibinfo
  {author} {\bibfnamefont {G.}~\bibnamefont {Kotliar}},\ and\ \bibinfo {author}
  {\bibfnamefont {A.}~\bibnamefont {Georges}},\ }\href
  {https://doi.org/10.1103/PhysRevB.80.064501} {\bibfield  {journal} {\bibinfo
  {journal} {Phys. Rev. B}\ }\textbf {\bibinfo {volume} {80}},\ \bibinfo
  {pages} {064501} (\bibinfo {year} {2009}{\natexlab{a}})}\BibitemShut
  {NoStop}%
\bibitem [{\citenamefont {Sakai}\ \emph {et~al.}(2009)\citenamefont {Sakai},
  \citenamefont {Motome},\ and\ \citenamefont {Imada}}]{sakai2009}%
  \BibitemOpen
  \bibfield  {author} {\bibinfo {author} {\bibfnamefont {S.}~\bibnamefont
  {Sakai}}, \bibinfo {author} {\bibfnamefont {Y.}~\bibnamefont {Motome}},\ and\
  \bibinfo {author} {\bibfnamefont {M.}~\bibnamefont {Imada}},\ }\href
  {https://doi.org/10.1103/PhysRevLett.102.056404} {\bibfield  {journal}
  {\bibinfo  {journal} {Phys. Rev. Lett.}\ }\textbf {\bibinfo {volume} {102}},\
  \bibinfo {pages} {056404} (\bibinfo {year} {2009})}\BibitemShut {NoStop}%
\bibitem [{\citenamefont {Ferrero}\ \emph
  {et~al.}(2009{\natexlab{b}})\citenamefont {Ferrero}, \citenamefont
  {Cornaglia}, \citenamefont {Leo}, \citenamefont {Parcollet}, \citenamefont
  {Kotliar},\ and\ \citenamefont {Georges}}]{ferrero2009epl}%
  \BibitemOpen
  \bibfield  {author} {\bibinfo {author} {\bibfnamefont {M.}~\bibnamefont
  {Ferrero}}, \bibinfo {author} {\bibfnamefont {P.~S.}\ \bibnamefont
  {Cornaglia}}, \bibinfo {author} {\bibfnamefont {L.~D.}\ \bibnamefont {Leo}},
  \bibinfo {author} {\bibfnamefont {O.}~\bibnamefont {Parcollet}}, \bibinfo
  {author} {\bibfnamefont {G.}~\bibnamefont {Kotliar}},\ and\ \bibinfo {author}
  {\bibfnamefont {A.}~\bibnamefont {Georges}},\ }\href
  {https://doi.org/10.1209/0295-5075/85/57009} {\bibfield  {journal} {\bibinfo
  {journal} {Europhys. Lett.}\ }\textbf {\bibinfo {volume} {85}},\ \bibinfo
  {pages} {57009} (\bibinfo {year} {2009}{\natexlab{b}})}\BibitemShut {NoStop}%
\bibitem [{\citenamefont {Sordi}\ \emph
  {et~al.}(2012{\natexlab{a}})\citenamefont {Sordi}, \citenamefont {S\'emon},
  \citenamefont {Haule},\ and\ \citenamefont {Tremblay}}]{sordi2012}%
  \BibitemOpen
  \bibfield  {author} {\bibinfo {author} {\bibfnamefont {G.}~\bibnamefont
  {Sordi}}, \bibinfo {author} {\bibfnamefont {P.}~\bibnamefont {S\'emon}},
  \bibinfo {author} {\bibfnamefont {K.}~\bibnamefont {Haule}},\ and\ \bibinfo
  {author} {\bibfnamefont {A.-M.~S.}\ \bibnamefont {Tremblay}},\ }\href
  {https://doi.org/10.1103/PhysRevLett.108.216401} {\bibfield  {journal}
  {\bibinfo  {journal} {Phys. Rev. Lett.}\ }\textbf {\bibinfo {volume} {108}},\
  \bibinfo {pages} {216401} (\bibinfo {year} {2012}{\natexlab{a}})}\BibitemShut
  {NoStop}%
\bibitem [{\citenamefont {Sordi}\ \emph
  {et~al.}(2012{\natexlab{b}})\citenamefont {Sordi}, \citenamefont {Sémon},
  \citenamefont {Haule},\ and\ \citenamefont {Tremblay}}]{sordi2012srep}%
  \BibitemOpen
  \bibfield  {author} {\bibinfo {author} {\bibfnamefont {G.}~\bibnamefont
  {Sordi}}, \bibinfo {author} {\bibfnamefont {P.}~\bibnamefont {Sémon}},
  \bibinfo {author} {\bibfnamefont {K.}~\bibnamefont {Haule}},\ and\ \bibinfo
  {author} {\bibfnamefont {A.-M.~S.}\ \bibnamefont {Tremblay}},\ }\href
  {https://doi.org/10.1038/srep00547} {\bibfield  {journal} {\bibinfo
  {journal} {Sci. Rep.}\ }\textbf {\bibinfo {volume} {2}},\ \bibinfo {pages}
  {547} (\bibinfo {year} {2012}{\natexlab{b}})}\BibitemShut {NoStop}%
\bibitem [{\citenamefont {Gull}\ \emph {et~al.}(2013)\citenamefont {Gull},
  \citenamefont {Parcollet},\ and\ \citenamefont {Millis}}]{gull2013}%
  \BibitemOpen
  \bibfield  {author} {\bibinfo {author} {\bibfnamefont {E.}~\bibnamefont
  {Gull}}, \bibinfo {author} {\bibfnamefont {O.}~\bibnamefont {Parcollet}},\
  and\ \bibinfo {author} {\bibfnamefont {A.~J.}\ \bibnamefont {Millis}},\
  }\href {https://doi.org/10.1103/PhysRevLett.110.216405} {\bibfield  {journal}
  {\bibinfo  {journal} {Phys. Rev. Lett.}\ }\textbf {\bibinfo {volume} {110}},\
  \bibinfo {pages} {216405} (\bibinfo {year} {2013})}\BibitemShut {NoStop}%
\bibitem [{\citenamefont {Kohno}(2014)}]{kohno2014}%
  \BibitemOpen
  \bibfield  {author} {\bibinfo {author} {\bibfnamefont {M.}~\bibnamefont
  {Kohno}},\ }\href {https://doi.org/10.1103/PhysRevB.90.035111} {\bibfield
  {journal} {\bibinfo  {journal} {Phys. Rev. B}\ }\textbf {\bibinfo {volume}
  {90}},\ \bibinfo {pages} {035111} (\bibinfo {year} {2014})}\BibitemShut
  {NoStop}%
\bibitem [{\citenamefont {Gunnarsson}\ \emph {et~al.}(2015)\citenamefont
  {Gunnarsson}, \citenamefont {Sch\"afer}, \citenamefont {LeBlanc},
  \citenamefont {Gull}, \citenamefont {Merino}, \citenamefont {Sangiovanni},
  \citenamefont {Rohringer},\ and\ \citenamefont {Toschi}}]{gunnarsson2015}%
  \BibitemOpen
  \bibfield  {author} {\bibinfo {author} {\bibfnamefont {O.}~\bibnamefont
  {Gunnarsson}}, \bibinfo {author} {\bibfnamefont {T.}~\bibnamefont
  {Sch\"afer}}, \bibinfo {author} {\bibfnamefont {J.~P.~F.}\ \bibnamefont
  {LeBlanc}}, \bibinfo {author} {\bibfnamefont {E.}~\bibnamefont {Gull}},
  \bibinfo {author} {\bibfnamefont {J.}~\bibnamefont {Merino}}, \bibinfo
  {author} {\bibfnamefont {G.}~\bibnamefont {Sangiovanni}}, \bibinfo {author}
  {\bibfnamefont {G.}~\bibnamefont {Rohringer}},\ and\ \bibinfo {author}
  {\bibfnamefont {A.}~\bibnamefont {Toschi}},\ }\href
  {https://doi.org/10.1103/PhysRevLett.114.236402} {\bibfield  {journal}
  {\bibinfo  {journal} {Phys. Rev. Lett.}\ }\textbf {\bibinfo {volume} {114}},\
  \bibinfo {pages} {236402} (\bibinfo {year} {2015})}\BibitemShut {NoStop}%
\bibitem [{\citenamefont {Yang}\ and\ \citenamefont
  {Feiguin}(2016)}]{yang2016}%
  \BibitemOpen
  \bibfield  {author} {\bibinfo {author} {\bibfnamefont {C.}~\bibnamefont
  {Yang}}\ and\ \bibinfo {author} {\bibfnamefont {A.~E.}\ \bibnamefont
  {Feiguin}},\ }\href {https://doi.org/10.1103/PhysRevB.93.081107} {\bibfield
  {journal} {\bibinfo  {journal} {Phys. Rev. B}\ }\textbf {\bibinfo {volume}
  {93}},\ \bibinfo {pages} {081107} (\bibinfo {year} {2016})}\BibitemShut
  {NoStop}%
\bibitem [{\citenamefont {Chen}\ \emph {et~al.}(2017)\citenamefont {Chen},
  \citenamefont {LeBlanc},\ and\ \citenamefont {Gull}}]{chen2017}%
  \BibitemOpen
  \bibfield  {author} {\bibinfo {author} {\bibfnamefont {X.}~\bibnamefont
  {Chen}}, \bibinfo {author} {\bibfnamefont {J.}~\bibnamefont {LeBlanc}},\ and\
  \bibinfo {author} {\bibfnamefont {E.}~\bibnamefont {Gull}},\ }\href
  {https://doi.org/10.1038/ncomms14986} {\bibfield  {journal} {\bibinfo
  {journal} {Nat. Commun.}\ }\textbf {\bibinfo {volume} {8}},\ \bibinfo {pages}
  {14986} (\bibinfo {year} {2017})}\BibitemShut {NoStop}%
\bibitem [{\citenamefont {Wu}\ \emph {et~al.}(2017)\citenamefont {Wu},
  \citenamefont {Ferrero}, \citenamefont {Georges},\ and\ \citenamefont
  {Kozik}}]{wu2017}%
  \BibitemOpen
  \bibfield  {author} {\bibinfo {author} {\bibfnamefont {W.}~\bibnamefont
  {Wu}}, \bibinfo {author} {\bibfnamefont {M.}~\bibnamefont {Ferrero}},
  \bibinfo {author} {\bibfnamefont {A.}~\bibnamefont {Georges}},\ and\ \bibinfo
  {author} {\bibfnamefont {E.}~\bibnamefont {Kozik}},\ }\href
  {https://doi.org/10.1103/PhysRevB.96.041105} {\bibfield  {journal} {\bibinfo
  {journal} {Phys. Rev. B}\ }\textbf {\bibinfo {volume} {96}},\ \bibinfo
  {pages} {041105} (\bibinfo {year} {2017})}\BibitemShut {NoStop}%
\bibitem [{\citenamefont {Wu}\ \emph {et~al.}(2018)\citenamefont {Wu},
  \citenamefont {Scheurer}, \citenamefont {Chatterjee}, \citenamefont
  {Sachdev}, \citenamefont {Georges},\ and\ \citenamefont {Ferrero}}]{wu2018}%
  \BibitemOpen
  \bibfield  {author} {\bibinfo {author} {\bibfnamefont {W.}~\bibnamefont
  {Wu}}, \bibinfo {author} {\bibfnamefont {M.~S.}\ \bibnamefont {Scheurer}},
  \bibinfo {author} {\bibfnamefont {S.}~\bibnamefont {Chatterjee}}, \bibinfo
  {author} {\bibfnamefont {S.}~\bibnamefont {Sachdev}}, \bibinfo {author}
  {\bibfnamefont {A.}~\bibnamefont {Georges}},\ and\ \bibinfo {author}
  {\bibfnamefont {M.}~\bibnamefont {Ferrero}},\ }\href
  {https://doi.org/10.1103/PhysRevX.8.021048} {\bibfield  {journal} {\bibinfo
  {journal} {Phys. Rev. X}\ }\textbf {\bibinfo {volume} {8}},\ \bibinfo {pages}
  {021048} (\bibinfo {year} {2018})}\BibitemShut {NoStop}%
\bibitem [{\citenamefont {Bragan\ifmmode~\mbox{\c{c}}\else \c{c}\fi{}a}\ \emph
  {et~al.}(2018)\citenamefont {Bragan\ifmmode~\mbox{\c{c}}\else \c{c}\fi{}a},
  \citenamefont {Sakai}, \citenamefont {Aguiar},\ and\ \citenamefont
  {Civelli}}]{braganca2018}%
  \BibitemOpen
  \bibfield  {author} {\bibinfo {author} {\bibfnamefont {H.}~\bibnamefont
  {Bragan\ifmmode~\mbox{\c{c}}\else \c{c}\fi{}a}}, \bibinfo {author}
  {\bibfnamefont {S.}~\bibnamefont {Sakai}}, \bibinfo {author} {\bibfnamefont
  {M.~C.~O.}\ \bibnamefont {Aguiar}},\ and\ \bibinfo {author} {\bibfnamefont
  {M.}~\bibnamefont {Civelli}},\ }\href
  {https://doi.org/10.1103/PhysRevLett.120.067002} {\bibfield  {journal}
  {\bibinfo  {journal} {Phys. Rev. Lett.}\ }\textbf {\bibinfo {volume} {120}},\
  \bibinfo {pages} {067002} (\bibinfo {year} {2018})}\BibitemShut {NoStop}%
\bibitem [{\citenamefont {Kuz'min}\ \emph {et~al.}(2020)\citenamefont
  {Kuz'min}, \citenamefont {Visotin}, \citenamefont {Nikolaev},\ and\
  \citenamefont {Ovchinnikov}}]{kuzmin2020}%
  \BibitemOpen
  \bibfield  {author} {\bibinfo {author} {\bibfnamefont {V.~I.}\ \bibnamefont
  {Kuz'min}}, \bibinfo {author} {\bibfnamefont {M.~A.}\ \bibnamefont
  {Visotin}}, \bibinfo {author} {\bibfnamefont {S.~V.}\ \bibnamefont
  {Nikolaev}},\ and\ \bibinfo {author} {\bibfnamefont {S.~G.}\ \bibnamefont
  {Ovchinnikov}},\ }\href {https://doi.org/10.1103/PhysRevB.101.115141}
  {\bibfield  {journal} {\bibinfo  {journal} {Phys. Rev. B}\ }\textbf {\bibinfo
  {volume} {101}},\ \bibinfo {pages} {115141} (\bibinfo {year}
  {2020})}\BibitemShut {NoStop}%
\bibitem [{\citenamefont {Moreo}(1993)}]{moreo1993}%
  \BibitemOpen
  \bibfield  {author} {\bibinfo {author} {\bibfnamefont {A.}~\bibnamefont
  {Moreo}},\ }\href {https://doi.org/10.1103/PhysRevB.48.3380} {\bibfield
  {journal} {\bibinfo  {journal} {Phys. Rev. B}\ }\textbf {\bibinfo {volume}
  {48}},\ \bibinfo {pages} {3380} (\bibinfo {year} {1993})}\BibitemShut
  {NoStop}%
\bibitem [{\citenamefont {Bon\ifmmode~\check{c}\else \v{c}\fi{}a}\ and\
  \citenamefont {Prelov\ifmmode~\check{s}\else
  \v{s}\fi{}ek}(2003)}]{bonca2003}%
  \BibitemOpen
  \bibfield  {author} {\bibinfo {author} {\bibfnamefont {J.}~\bibnamefont
  {Bon\ifmmode~\check{c}\else \v{c}\fi{}a}}\ and\ \bibinfo {author}
  {\bibfnamefont {P.}~\bibnamefont {Prelov\ifmmode~\check{s}\else
  \v{s}\fi{}ek}},\ }\href {https://doi.org/10.1103/PhysRevB.67.085103}
  {\bibfield  {journal} {\bibinfo  {journal} {Phys. Rev. B}\ }\textbf {\bibinfo
  {volume} {67}},\ \bibinfo {pages} {085103} (\bibinfo {year}
  {2003})}\BibitemShut {NoStop}%
\bibitem [{\citenamefont {Kokalj}(2017)}]{kokalj2017}%
  \BibitemOpen
  \bibfield  {author} {\bibinfo {author} {\bibfnamefont {J.}~\bibnamefont
  {Kokalj}},\ }\href {https://doi.org/10.1103/PhysRevB.95.041110} {\bibfield
  {journal} {\bibinfo  {journal} {Phys. Rev. B}\ }\textbf {\bibinfo {volume}
  {95}},\ \bibinfo {pages} {041110} (\bibinfo {year} {2017})}\BibitemShut
  {NoStop}%
\bibitem [{\citenamefont {Reymbaut}\ \emph {et~al.}(2019)\citenamefont
  {Reymbaut}, \citenamefont {Bergeron}, \citenamefont {Garioud}, \citenamefont
  {Th\'enault}, \citenamefont {Charlebois}, \citenamefont {S\'emon},\ and\
  \citenamefont {Tremblay}}]{reymbaut2019}%
  \BibitemOpen
  \bibfield  {author} {\bibinfo {author} {\bibfnamefont {A.}~\bibnamefont
  {Reymbaut}}, \bibinfo {author} {\bibfnamefont {S.}~\bibnamefont {Bergeron}},
  \bibinfo {author} {\bibfnamefont {R.}~\bibnamefont {Garioud}}, \bibinfo
  {author} {\bibfnamefont {M.}~\bibnamefont {Th\'enault}}, \bibinfo {author}
  {\bibfnamefont {M.}~\bibnamefont {Charlebois}}, \bibinfo {author}
  {\bibfnamefont {P.}~\bibnamefont {S\'emon}},\ and\ \bibinfo {author}
  {\bibfnamefont {A.-M.~S.}\ \bibnamefont {Tremblay}},\ }\href
  {https://doi.org/10.1103/PhysRevResearch.1.023015} {\bibfield  {journal}
  {\bibinfo  {journal} {Phys. Rev. Research}\ }\textbf {\bibinfo {volume}
  {1}},\ \bibinfo {pages} {023015} (\bibinfo {year} {2019})}\BibitemShut
  {NoStop}%
\bibitem [{\citenamefont {White}\ and\ \citenamefont
  {Scalapino}(2003)}]{white2003}%
  \BibitemOpen
  \bibfield  {author} {\bibinfo {author} {\bibfnamefont {S.~R.}\ \bibnamefont
  {White}}\ and\ \bibinfo {author} {\bibfnamefont {D.~J.}\ \bibnamefont
  {Scalapino}},\ }\href {https://doi.org/10.1103/PhysRevLett.91.136403}
  {\bibfield  {journal} {\bibinfo  {journal} {Phys. Rev. Lett.}\ }\textbf
  {\bibinfo {volume} {91}},\ \bibinfo {pages} {136403} (\bibinfo {year}
  {2003})}\BibitemShut {NoStop}%
\bibitem [{\citenamefont {Zheng}\ \emph {et~al.}(2017)\citenamefont {Zheng},
  \citenamefont {Chung}, \citenamefont {Corboz}, \citenamefont {Ehlers},
  \citenamefont {Qin}, \citenamefont {Noack}, \citenamefont {Shi},
  \citenamefont {White}, \citenamefont {Zhang},\ and\ \citenamefont
  {Chan}}]{zheng2017}%
  \BibitemOpen
  \bibfield  {author} {\bibinfo {author} {\bibfnamefont {B.-X.}\ \bibnamefont
  {Zheng}}, \bibinfo {author} {\bibfnamefont {C.-M.}\ \bibnamefont {Chung}},
  \bibinfo {author} {\bibfnamefont {P.}~\bibnamefont {Corboz}}, \bibinfo
  {author} {\bibfnamefont {G.}~\bibnamefont {Ehlers}}, \bibinfo {author}
  {\bibfnamefont {M.-P.}\ \bibnamefont {Qin}}, \bibinfo {author} {\bibfnamefont
  {R.~M.}\ \bibnamefont {Noack}}, \bibinfo {author} {\bibfnamefont
  {H.}~\bibnamefont {Shi}}, \bibinfo {author} {\bibfnamefont {S.~R.}\
  \bibnamefont {White}}, \bibinfo {author} {\bibfnamefont {S.}~\bibnamefont
  {Zhang}},\ and\ \bibinfo {author} {\bibfnamefont {G.~K.-L.}\ \bibnamefont
  {Chan}},\ }\href {https://doi.org/10.1126/science.aam7127} {\bibfield
  {journal} {\bibinfo  {journal} {Science}\ }\textbf {\bibinfo {volume}
  {358}},\ \bibinfo {pages} {1155} (\bibinfo {year} {2017})}\BibitemShut
  {NoStop}%
\bibitem [{\citenamefont {Huang}\ \emph {et~al.}(2018)\citenamefont {Huang},
  \citenamefont {Mendl}, \citenamefont {Jiang}, \citenamefont {Moritz},\ and\
  \citenamefont {Devereaux}}]{huang2018}%
  \BibitemOpen
  \bibfield  {author} {\bibinfo {author} {\bibfnamefont {E.~W.}\ \bibnamefont
  {Huang}}, \bibinfo {author} {\bibfnamefont {C.~B.}\ \bibnamefont {Mendl}},
  \bibinfo {author} {\bibfnamefont {H.-C.}\ \bibnamefont {Jiang}}, \bibinfo
  {author} {\bibfnamefont {B.}~\bibnamefont {Moritz}},\ and\ \bibinfo {author}
  {\bibfnamefont {T.~P.}\ \bibnamefont {Devereaux}},\ }\href
  {https://doi.org/10.1038/s41535-018-0097-0} {\bibfield  {journal} {\bibinfo
  {journal} {npj Quant. Mat.}\ }\textbf {\bibinfo {volume} {3}},\ \bibinfo
  {pages} {22} (\bibinfo {year} {2018})}\BibitemShut {NoStop}%
\bibitem [{\citenamefont {Gros}(1996)}]{gros1996}%
  \BibitemOpen
  \bibfield  {author} {\bibinfo {author} {\bibfnamefont {C.}~\bibnamefont
  {Gros}},\ }\href {https://doi.org/10.1103/PhysRevB.53.6865} {\bibfield
  {journal} {\bibinfo  {journal} {Phys. Rev. B}\ }\textbf {\bibinfo {volume}
  {53}},\ \bibinfo {pages} {6865} (\bibinfo {year} {1996})}\BibitemShut
  {NoStop}%
\bibitem [{\citenamefont {Lin}\ \emph {et~al.}(2001)\citenamefont {Lin},
  \citenamefont {Zong},\ and\ \citenamefont {Ceperley}}]{lin2001}%
  \BibitemOpen
  \bibfield  {author} {\bibinfo {author} {\bibfnamefont {C.}~\bibnamefont
  {Lin}}, \bibinfo {author} {\bibfnamefont {F.~H.}\ \bibnamefont {Zong}},\ and\
  \bibinfo {author} {\bibfnamefont {D.~M.}\ \bibnamefont {Ceperley}},\ }\href
  {https://doi.org/10.1103/PhysRevE.64.016702} {\bibfield  {journal} {\bibinfo
  {journal} {Phys. Rev. E}\ }\textbf {\bibinfo {volume} {64}},\ \bibinfo
  {pages} {016702} (\bibinfo {year} {2001})}\BibitemShut {NoStop}%
\bibitem [{\citenamefont {Koretsune}\ \emph {et~al.}(2007)\citenamefont
  {Koretsune}, \citenamefont {Motome},\ and\ \citenamefont
  {Furusaki}}]{koretsune2007}%
  \BibitemOpen
  \bibfield  {author} {\bibinfo {author} {\bibfnamefont {T.}~\bibnamefont
  {Koretsune}}, \bibinfo {author} {\bibfnamefont {Y.}~\bibnamefont {Motome}},\
  and\ \bibinfo {author} {\bibfnamefont {A.}~\bibnamefont {Furusaki}},\ }\href
  {https://doi.org/10.1143/JPSJ.76.074719} {\bibfield  {journal} {\bibinfo
  {journal} {Journal of the Physical Society of Japan}\ }\textbf {\bibinfo
  {volume} {76}},\ \bibinfo {pages} {074719} (\bibinfo {year}
  {2007})}\BibitemShut {NoStop}%
\bibitem [{\citenamefont {Qin}\ \emph {et~al.}(2016)\citenamefont {Qin},
  \citenamefont {Shi},\ and\ \citenamefont {Zhang}}]{qin2016}%
  \BibitemOpen
  \bibfield  {author} {\bibinfo {author} {\bibfnamefont {M.}~\bibnamefont
  {Qin}}, \bibinfo {author} {\bibfnamefont {H.}~\bibnamefont {Shi}},\ and\
  \bibinfo {author} {\bibfnamefont {S.}~\bibnamefont {Zhang}},\ }\href
  {https://doi.org/10.1103/PhysRevB.94.085103} {\bibfield  {journal} {\bibinfo
  {journal} {Phys. Rev. B}\ }\textbf {\bibinfo {volume} {94}},\ \bibinfo
  {pages} {085103} (\bibinfo {year} {2016})}\BibitemShut {NoStop}%
\bibitem [{\citenamefont {Karakuzu}\ \emph {et~al.}(2017)\citenamefont
  {Karakuzu}, \citenamefont {Tocchio}, \citenamefont {Sorella},\ and\
  \citenamefont {Becca}}]{karakuzu2017}%
  \BibitemOpen
  \bibfield  {author} {\bibinfo {author} {\bibfnamefont {S.}~\bibnamefont
  {Karakuzu}}, \bibinfo {author} {\bibfnamefont {L.~F.}\ \bibnamefont
  {Tocchio}}, \bibinfo {author} {\bibfnamefont {S.}~\bibnamefont {Sorella}},\
  and\ \bibinfo {author} {\bibfnamefont {F.}~\bibnamefont {Becca}},\ }\href
  {https://doi.org/10.1103/PhysRevB.96.205145} {\bibfield  {journal} {\bibinfo
  {journal} {Phys. Rev. B}\ }\textbf {\bibinfo {volume} {96}},\ \bibinfo
  {pages} {205145} (\bibinfo {year} {2017})}\BibitemShut {NoStop}%
\bibitem [{\citenamefont {Karakuzu}\ \emph {et~al.}(2018)\citenamefont
  {Karakuzu}, \citenamefont {Seki},\ and\ \citenamefont
  {Sorella}}]{karakuzu2018}%
  \BibitemOpen
  \bibfield  {author} {\bibinfo {author} {\bibfnamefont {S.}~\bibnamefont
  {Karakuzu}}, \bibinfo {author} {\bibfnamefont {K.}~\bibnamefont {Seki}},\
  and\ \bibinfo {author} {\bibfnamefont {S.}~\bibnamefont {Sorella}},\ }\href
  {https://doi.org/10.1103/PhysRevB.98.075156} {\bibfield  {journal} {\bibinfo
  {journal} {Phys. Rev. B}\ }\textbf {\bibinfo {volume} {98}},\ \bibinfo
  {pages} {075156} (\bibinfo {year} {2018})}\BibitemShut {NoStop}%
\bibitem [{\citenamefont {Nagaoka}(1966)}]{nagaoka}%
  \BibitemOpen
  \bibfield  {author} {\bibinfo {author} {\bibfnamefont {Y.}~\bibnamefont
  {Nagaoka}},\ }\href {https://doi.org/10.1103/PhysRev.147.392} {\bibfield
  {journal} {\bibinfo  {journal} {Phys. Rev.}\ }\textbf {\bibinfo {volume}
  {147}},\ \bibinfo {pages} {392} (\bibinfo {year} {1966})}\BibitemShut
  {NoStop}%
\bibitem [{\citenamefont {Buterakos}\ and\ \citenamefont
  {Das~Sarma}(2019)}]{buterakos2019}%
  \BibitemOpen
  \bibfield  {author} {\bibinfo {author} {\bibfnamefont {D.}~\bibnamefont
  {Buterakos}}\ and\ \bibinfo {author} {\bibfnamefont {S.}~\bibnamefont
  {Das~Sarma}},\ }\href {https://doi.org/10.1103/PhysRevB.100.224421}
  {\bibfield  {journal} {\bibinfo  {journal} {Phys. Rev. B}\ }\textbf {\bibinfo
  {volume} {100}},\ \bibinfo {pages} {224421} (\bibinfo {year}
  {2019})}\BibitemShut {NoStop}%
\bibitem [{\citenamefont {Gr\"ober}\ \emph {et~al.}(2000)\citenamefont
  {Gr\"ober}, \citenamefont {Eder},\ and\ \citenamefont {Hanke}}]{grober2000}%
  \BibitemOpen
  \bibfield  {author} {\bibinfo {author} {\bibfnamefont {C.}~\bibnamefont
  {Gr\"ober}}, \bibinfo {author} {\bibfnamefont {R.}~\bibnamefont {Eder}},\
  and\ \bibinfo {author} {\bibfnamefont {W.}~\bibnamefont {Hanke}},\ }\href
  {https://doi.org/10.1103/PhysRevB.62.4336} {\bibfield  {journal} {\bibinfo
  {journal} {Phys. Rev. B}\ }\textbf {\bibinfo {volume} {62}},\ \bibinfo
  {pages} {4336} (\bibinfo {year} {2000})}\BibitemShut {NoStop}%
\end{thebibliography}
%

\end{document}